\documentclass[preprint,showpacs,preprintnumbers,amsmath,amssymb]{revtex4}

\usepackage{graphicx}
\usepackage{dcolumn}
\usepackage{bm}

\begin{document}

\noindent

\preprint{}

\title{Madelung Fluid Model for The Most Likely Wave Function\\ of a Single Free Particle in Two Dimensional Space\\ with a Given Average Energy}

\author{Agung Budiyono and Ken Umeno}

\affiliation{Institute for the Physical and Chemical Research (RIKEN), 2-1 Hirosawa, Wako-shi, Saitama 351-0198, Japan}

\date{\today}

\begin{abstract} 

We consider spatially two dimensional Madelung fluid whose irrotational motion reduces into the Schr\"odinger equation for a single free particle. In this respect, we regard the former as a direct generalization of the latter, allowing a rotational quantum flow. We then ask for the most likely wave function possessing a given average energy by maximizing the Shannon information entropy over the quantum probability density. We show that there exists a class of solutions in which the wave function is self-trapped, rotationally symmetric, spatially localized with finite support, and spinning around its center, yet stationary. The stationarity comes from the balance between the attractive quantum force field of a trapping quantum potential generated by quantum probability density and the repulsive centrifugal force of a rotating velocity vector field. We further show that there is a limiting case where the wave function is non-spinning and yet still stationary. This special state turns out to be the lowest stationary state of the ordinary Schr\"odinger equation for a particle in a cylindrical tube classical potential. 

\end{abstract}

\maketitle

\section{Madelung Fluid: Generalized Schr\"odinger Equation}

Let us consider a quantum system of a single free particle with mass $m$ whose dynamics is confined in two dimensional space, ${\bf q}=\{x,y\}$. The complex-valued wave function $\psi({\bf q};t)$, where $t$ is time, is governed by the linear Schr\"odinger equation 
\begin{equation}
i\hbar\partial_t\psi({\bf q};t)=-\frac{\hbar^2}{2m}\partial_q^2\psi({\bf q};t),
\label{Schroedinger equation}
\end{equation}
where $\partial_q^2=\partial_{\bf q}\cdot\partial_{\bf q}$ is two dimensional Laplace operator. 

Let us project the above dynamics onto real-space. To do this, let us write the wave function in polar form, $\psi({\bf q};t)=R({\bf q};t)\exp(iS({\bf q};t)/\hbar)$, where the quantum amplitude $R$ and the quantum phase $S$ are real-valued functions. Inserting this into Eq. (\ref{Schroedinger equation}) and separating into the real and imaginary parts, one gets
\begin{eqnarray}
\partial_tS+\frac{\partial_{\bf q}S\cdot\partial_{\bf q}S}{2m}+U=0,\hspace{10mm}\nonumber\\
\partial_t\rho+\partial_{\bf q}\cdot\Big(\frac{\partial_{\bf q}S}{m}\rho\Big)=0,\hspace{15mm}
\label{HJ-C equation}
\end{eqnarray}
where $U$ is the so-called quantum potential generated by the quantum probability density $\rho=\psi\psi^*=R^2$ as 
\begin{equation}
U({\bf q};t)=-\frac{\hbar^2}{2m}\frac{\partial_q^2R}{R}.
\label{quantum potential}
\end{equation}

Next, let us postulate a velocity flow generated by the quantum phase as follows\begin{equation}
{\bf v}({\bf q};t)=\frac{1}{m}\partial_{\bf q}S({\bf q};t). 
\label{velocity field-quantum phase}
\end{equation}
Using this, the coupled equations in (\ref{HJ-C equation}) can be rewritten into the following form
\begin{eqnarray}
m\frac{d{\bf v}}{dt}=-\partial_{\bf q}U,\hspace{5mm}\nonumber\\
\partial_t\rho+\partial_{\bf q}\cdot({\bf v}\rho)=0.
\label{Madelung fluid dynamics}
\end{eqnarray}
The upper equation can then be seen as an Euler equation for the velocity flow ${\bf v}({\bf q};t)$ dragged by a special kind of force field generated by the quantum amplitude, $-\partial_{\bf q}U$, later on to be referred to as quantum force. On the other hand, the lower equation can be regarded as the continuity equation for the quantum probability density, $\rho({\bf q};t)$. We have thus a non-linear fluid dynamics in real-space. This real-space hydrodynamical interpretation of Schr\"odinger equation is introduced by Madelung as soon as 1926 \cite{Madelung paper}. Moreover, the same mathematical formalism is also used to based what is later called as pilot-wave interpretation, initiated by de Broglie and expanded by Bohm \cite{de Broglie papers,Bohm paper,Bohm-Hiley's book,Holland's book}. 

If we assume that the quantum phase $S$ is twice differentiable, then the velocity flow defined in Eq. (\ref{velocity field-quantum phase}) is irrotational 
\begin{equation}
\partial_{\bf q}\times {\bf v}=\frac{1}{m}\Big(\frac{\partial^2S}{\partial_x\partial_y}-\frac{\partial^2 S}{\partial_y\partial_x}\Big){\bf \hat{z}}={\bf 0},
\label{irrotational flow}
\end{equation} 
where ${\bf \hat{z}}$ is a unit vector orthogonal to the $xy-$plane. In this present work, we shall follow the suggestion made by Takabayashi \cite{Takabayashi paper} to consider the pair of equations in (\ref{Madelung fluid dynamics}) as a direct generalization of Schr\"odinger equation by allowing rotational flow. The special case of Madelung fluid with only irrotational flow will then reduce into the Schr\"odinger equation of (\ref{Schroedinger equation}). In other words, we shall consider Eqs. (\ref{Madelung fluid dynamics}) as describing the quantum system of a single free particle, allowing a class of wave functions whose quantum phase possesses singularity such that Eq. (\ref{irrotational flow}) is no more valid. 

One of the important feature of the above Madelung fluid dynamics is that it is  self-referential. One can see that the quantum probability density $\rho({\bf q};t)$ will generate quantum potential $U({\bf q};t)$ through Eq. (\ref{quantum potential}), and in turn $U({\bf q};t)$ will tell $\rho({\bf q};t)$ the way it must evolve through Eqs. (\ref{Madelung fluid dynamics}), and so on and so forth. There is a dynamics circularity between the object to be ruled, $\rho$, and the rule, $U$. Self-referential property is a sympton of complex non-linear systems, and is argued to be the general origin for the emergence of many interesting phenomena observed in Nature \cite{self-referential dynamics and complexity}. Hence, it is of great interest to evaluate the fixed point of the self-referential Madelung fluid dynamics. 

In this paper, we would like to address the following simple yet fundamental question: What is the most likely wave function of a single free particle with the average quantum mechanical energy $\langle E\rangle$ given by
\begin{equation}
\langle E\rangle=\int dq\hspace{1mm}\psi^*({\bf q})\Big(-\frac{\hbar^2}{2m}\partial_q^2\Big)\psi({\bf q}),
\label{average quantum mechanical energy}
\end{equation}
where $dq=dxdy$. We shall show that there exist a class of solutions which turns out to be the fixed point of the self-referential Madelung fluid dynamics described above. 

\section{A Class of Self-Trapped Quantum Probability Densities}

\begin{figure}[tbp]
\begin{center}
\includegraphics*[width=10cm]{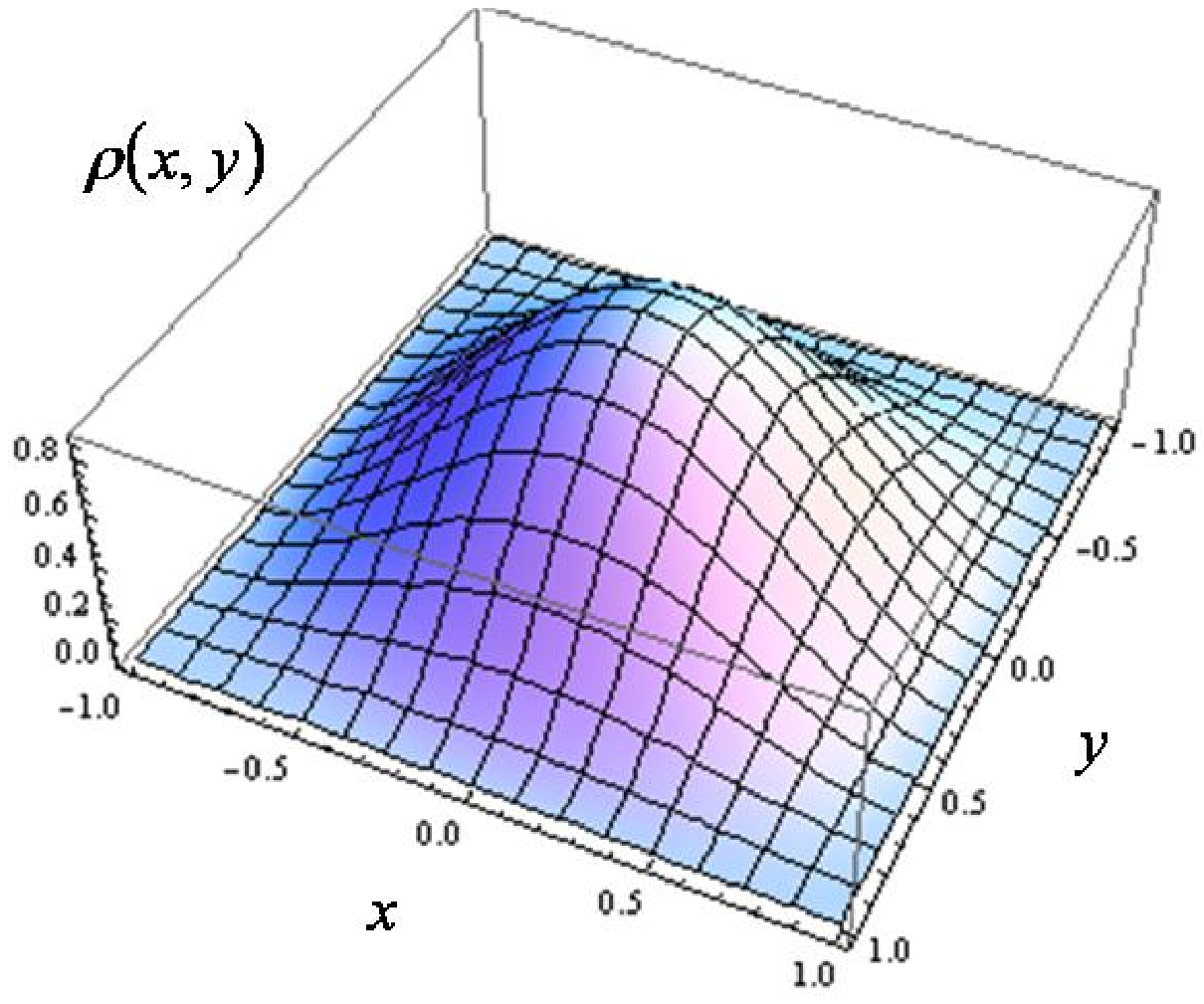}
\includegraphics*[width=10cm]{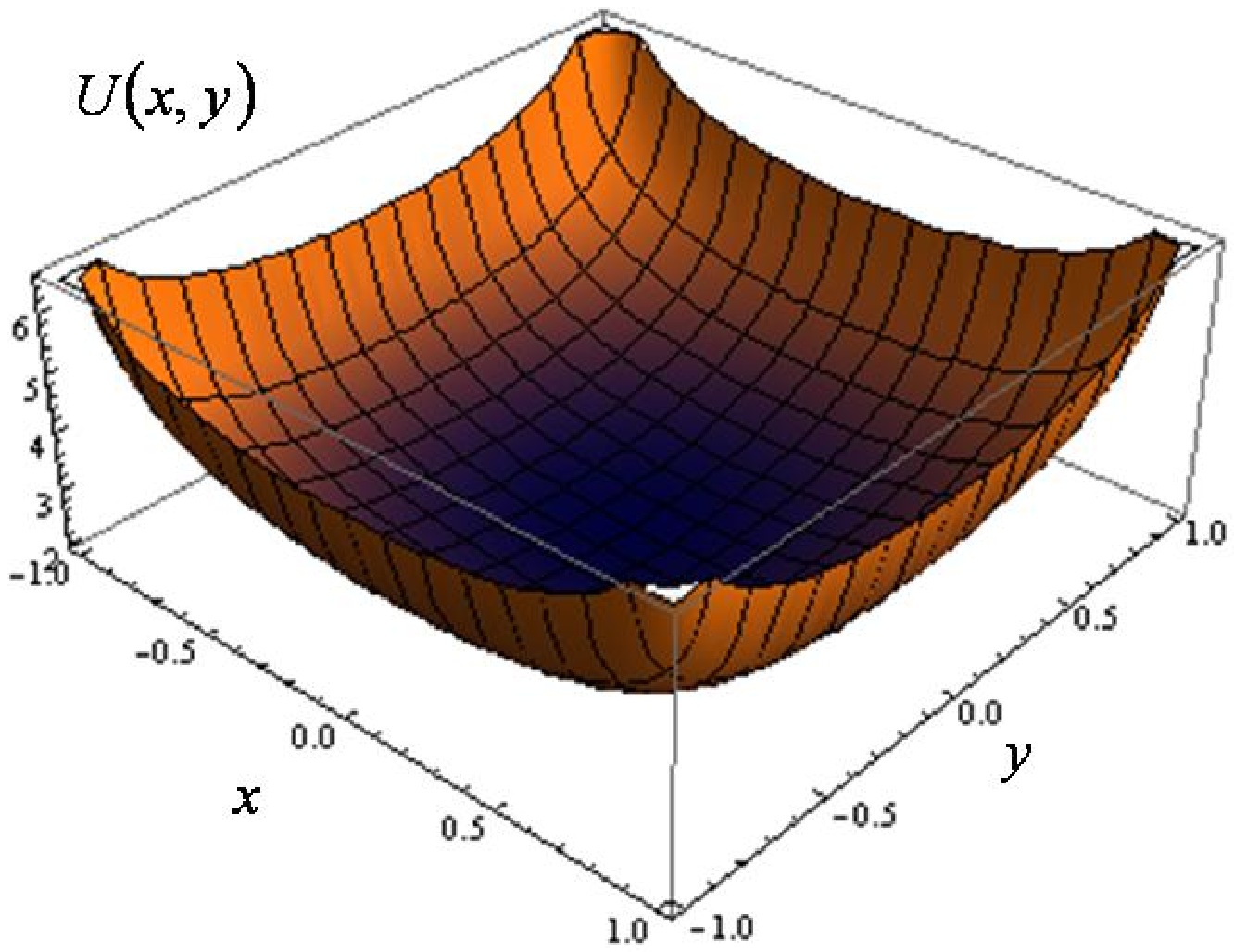}
\end{center}
\caption{The profile of a self-trapped quantum probability density (upper) and the corresponding quantum potential it generates (lower).}
\label{2D QP-QPD}
\end{figure}

To investigate the question set up at the end of the previous section, let us discuss the following readily executable problem. First, in quantum mechanics $\rho({\bf{q}})$ provides the essential information about the position of the particle \cite{Copenhagen-ontological interpretation,Isham's book,Bohm-Hiley's book, Bell's unspeakable}. Let us quantify this information using differential entropy or Shannon information entropy for continuous random variable as \cite{Shannon entropy,Cover's book}
\begin{equation}
H[\rho]=-\int dq\hspace{1mm}\rho({\bf q})\ln\rho({\bf q}).
\label{Shannon entropy}
\end{equation}
It measures the degree of spatial localization of the quantum probability density, $\rho({\bf q})$. Now, let us search for a class of wave functions which maximizes the above Shannon information entropy, given a finite value of the average quantum potential
\begin{equation}
{\bar U}=\int dq\hspace{1mm}U({\bf q})\rho({\bf q}).
\label{average quantum potential}
\end{equation} 

This is the so-called maximum entropy principle \cite{Jaynes-MEP}. It has been argued as the only way to infer from an incomplete information which does not lead to logical inconsistency \cite{Shore-Johnson-MEP}. Hence, it will give us the most likely $\rho({\bf q})$ whose average quantum potential is equal to $\bar{U}$. Maximizing Eq. (\ref{Shannon entropy}) with constraint of Eq. (\ref{average quantum potential}), one directly obtains the following relation \cite{Mackey-MEP}
\begin{equation}
\rho({\bf q};t)=\frac{1}{Z(\beta)}\exp\big(-\beta U({\bf q};t)\big), 
\label{canonical quantum probability density}  
\end{equation}
where $\beta$ is a real-valued constant (Lagrange constant), and $Z$ is a normalization factor given by 
\begin{equation}
Z=\int dq \exp(-\beta U).
\label{normalization constant} 
\end{equation}
Notice that Eq. (\ref{canonical quantum probability density}) together with the definition of quantum potential given in Eq. (\ref{quantum potential}) comprise a differential equation for $\rho({\bf q})$ or $U({\bf q})$, subjected to the condition that $\rho({\bf q})$ must be normalized. $\beta$ can then be calculated by inserting Eq. (\ref{canonical quantum probability density}) back into Eq. (\ref{average quantum potential}). Hence, it is a function of the average quantum potential, $\beta=\beta(\bar{U})$. Below, instead of using $\bar{U}$, we shall consider $\beta$ as parameter of the class of quantum probability density $\rho({\bf q};\beta)$ maximizing Shannon information entropy. Furthermore, for later convenient, we shall limit ourselves to positive definite $\beta$.

Let us show that there are infinitely many quantum probability densities satisfying Eq. (\ref{canonical quantum probability density}). To do this, applying Laplace operator to both sides of Eq. (\ref{canonical quantum probability density}), using the fact that $\rho=R^2$ and the definition of quantum potential given in Eq. (\ref{quantum potential}) one obtains the following non-linear partial differential equation 
\begin{equation}
\partial_q^2U=\frac{\beta}{2}\partial_{\bf q}U\cdot\partial_{\bf q}U+\frac{4m}{\hbar^2\beta}U.\label{NPDE for quantum potential}
\end{equation}
Let us remark first that the above differential equation is invariant under the following rotation of coordinate system  \begin{eqnarray}
x'=x\cos\theta-y\sin\theta,\nonumber\\
y'=x\sin\theta+y\cos\theta,
\label{rotation}
\end{eqnarray}
where $\theta$ is the angle of rotation. Hence, if $U({\bf q})$ is a solution then so is $U({\bf q}')$.

We shall resort to numerical methods to solve Eq. (\ref{NPDE for quantum potential}). For simplicity, let us confine ourselves to a class of solutions in which the quantum probability density is separable
\begin{equation}
\rho({\bf q})=\rho_x(x)\rho_y(y),
\label{separable quantum probability density}
\end{equation}
where $\rho_i(i)$ depends only on $i=(x,y)$. In this case, the quantum amplitude is also separable $R({\bf q})=R_x(x)R_y(y)$ such that the quantum potential is decomposable as
\begin{equation}
U({\bf q})=U_x(x)+U_y(y),
\label{decomposible quantum potential}
\end{equation}
where each term on the right hand side is given by
\begin{equation}
U_i(i)=-\frac{\hbar^2}{2m}\frac{\partial_i^2R_i}{R_i},\hspace{2mm}i=x,y.
\label{partial quantum potential}
\end{equation}

Using Eq. (\ref{decomposible quantum potential}), Eq. (\ref{NPDE for quantum potential}) can be collected as
\begin{equation}
\partial_x^2U_x-\frac{\beta}{2}(\partial_xU_x)^2-\Lambda^2U_x=-\partial_y^2U_y+\frac{\beta}{2}(\partial_yU_y)^2+\Lambda^2 U_y.
\label{separation of variable 1}
\end{equation}
Hence, both sides must be equal to a constant, say $E$. Let us choose the case when $E=0$. One therefore obtains the following two de-coupled ordinary differential equations
\begin{eqnarray}
\partial_x^2U_x=\frac{\beta}{2}(\partial_xU_x)^2+\Lambda^2U_x,\nonumber\\
\partial_y^2U_y=\frac{\beta}{2}(\partial_yU_y)^2+\Lambda^2 U_y.
\label{un-coupled differential equation}
\end{eqnarray}
Fig. \ref{2D QP-QPD} shows the numerical solution of Eqs. (\ref{un-coupled differential equation}) with the boundary conditions $\partial_xU_x(0)=\partial_yU_y(0)=0$ and $U_x(0)=U_y(0)=1$, for $\beta=1$. For convenient, all numerical calculations in this paper is done by putting $m=\hbar=1$. The profile of quantum probability density (upper) is plotted together with the profile of quantum potential it itself generates (lower). We can see clearly from the numerical solution that globally the quantum probability density is being trapped by its own quantum potential. 

The global self-trapping property can be justified for any positive value of $U_i(0)>0$ as follows. First, from the upper part of Eq. (\ref{un-coupled differential equation}), applying the boundary conditions at $x=0$, one gets $\partial_x^2U_x(0)=\Lambda U_x(0) > 0$, such that $U_x(x)$ is locally convex at $x=0$. Hence, since $\partial_xU_x(0)=0$, at spatial points nearby $x=0$ one has $U_x(x)>U_x(0)>0$. Moreover, since the first term on the right hand side is always non-negative, at this region one has $\partial_x^2U_x>0$. This geometrical reasoning can be extended such that for the whole spatial points one gets
\begin{equation}
\partial_x^2U_x>0. 
\label{convex quantum potential}
\end{equation}
The same thing applies for $U_y$. On the other hand, from Eq. (\ref{decomposible quantum potential}) one has $\partial_x\partial_yU=0$. In general one therefore obtains $\partial_i\partial_jU\geq 0$. Hence, the case when $U_i(0)> 0$ will give an everywhere positive and convex quantum potential. ${\bf q}={\bf 0}$ turns out to be the global minimum of $U({\bf q})$. One can thus conclude that quantum probability density, $\rho({\bf q})$, is being trapped by its own self-generated quantum potential, $U({\bf q})$. Moreover, the rotational invariant of the differential equation (\ref{NPDE for quantum potential}) guarantees that one can generate a new solution by rotating the solution given in Fig. \ref{2D QP-QPD}. We have thus a class of infinitely many self-trapped quantum probability densities characterized by $\beta$. 

Next, using the fact that $U_i(i)$ is positive everywhere and Eq. (\ref{partial quantum potential}), one gets 
\begin{equation}
\partial_i^2R_i \le 0,\hspace{2mm}i=(x,y).
\label{concave quantum amplitude}
\end{equation} 
Hence, $R_i(i)$ is everywhere concave. Since $R_i(i)$ is finite and possesses symmetrical property $R_i(i)=R_i(-i)$, then $R_i(i)$ must cross the $i-$axis at finite value of $i=\pm i_m$. In other words, $R_i(i)$ possesses only finite support on $i-$axis, namely $[-i_m,i_m]$. One can thus conclude that $R({\bf q})$ has only a finite support on $xy-$plane, that is all points ${\bf q}$ belonging to the rectangle $[-x_m,x_m]\otimes[-y_m,y_m]$. At the boundary of this rectangle, namely the lines $i=\pm i_m$, $i=(x,y)$, $U(x,y)$ is infinite, yet $R(x,y)$ is vanishing. We shall show later that for a subclass of solutions, $\bar{U}$ remains finite for non-vanishing $\beta$. Moreover, at $i=\pm i_m$, $\partial_iR_i$ is not continuous but $\partial_i\rho_i=2R_i\partial_iR_i$ is continuous and equal to zero.

We have thus developed a class of quantum probability densities satisfying Eq. (\ref{canonical quantum probability density}) which is self-trapped by its own self-generated quantum potential. Yet, we have not specified the quantum phase or the velocity vector field, to completely identify our Madelung fluid dynamics. Nevertheless, one can see that due to the global trapping property of the quantum potential, any initial velocity vector field will be dragged by the quantum force to localize even further. In the next section, we shall specify a specific velocity vector field which generates a force that exactly cancels the attractive quantum force of the trapping quantum potential to create a stationary Madelung flow.

\section{A Class of Stationary-Spinning Wave Functions}

\begin{figure}[tbp]
\begin{center}
\includegraphics*[width=10cm]{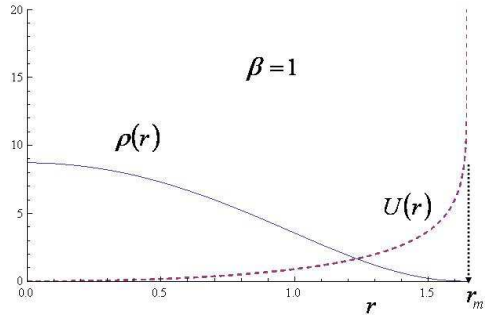}
\end{center}
\caption{The radial profile of a self-trapped quantum probability density (solid line) trapped by its own quantum potential (dashed line). The quantum potential is shifted down such that its minimum is equal to zero and the quantum probability density is rescaled by a constant factor. See text for detail.}
\label{rotationally symmetric self-trapped QPD}
\end{figure}

Let us find a class of self-trapped wave functions satisfying the differential equation (\ref{NPDE for quantum potential}) which is rotationally symmetric. They are the eigenfunctions of the rotation operator. To do this, it is convenient to use polar coordinate $\{r,\theta\}$, where $r=\sqrt{x^2+y^2}$ and $\theta=\tan^{-1}(y/x)$. For a rotationally symmetric solution, one imposes $\partial_{\theta}U=0$. This condition leads to $\{\partial_x,\partial_y\}=\{\cos\theta\partial_r,\sin\theta\partial_r\}$. Inserting these into Eq. (\ref{NPDE for quantum potential}), one finally has to solve
\begin{equation}
\partial_r^2U+\frac{2}{r}\partial_rU-\frac{\beta}{2}(\partial_rU)^2-\frac{4m}{\hbar^2\beta}U=0.
\label{NPDE for quantum potential-rotationally symmetric}
\end{equation}
Fig. \ref{rotationally symmetric self-trapped QPD} shows the numerical solutions of Eq. (\ref{NPDE for quantum potential-rotationally symmetric}) for $\beta=1$. $\rho(r)$ (solid line) is plotted together with the corresponding $U(r)$ (dashed line). We can see clearly that globally the quantum probability density is indeed being trapped by the quantum potential it itself generates. Moreover, for a rotationally symmetric quantum potential, the attractive quantum force vector field, $-\partial_{\bf q}U$, is directing toward the origin, ${\bf q}={\bf 0}$.

\begin{figure}[tbp]
\begin{center}
\includegraphics*[width=10cm]{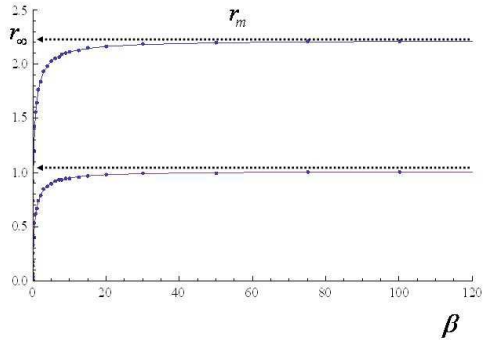}
\end{center}
\caption{$r_m$ (upper) and $\bar{r^2}$ (lower) against $\beta$.}
\label{radius of support}
\end{figure}

Numerical solutions of the differential equation (\ref{NPDE for quantum potential-rotationally symmetric}) show that again the quantum probability density possesses a finite support, $\mathcal{M}$. Since the quantum probability is rotationally symmetric, then the support is a disk with finite radius. Let us denotes the radius as $r=r_m$. At the boundary of the support, $\partial\mathcal{M}$, which makes a circle $x^2+y^2=r_m^2$, the quantum potential is infinite, $U(r_m)=\infty$. See Fig. \ref{rotationally symmetric self-trapped QPD}. In Fig. \ref{radius of support}, we plot the radius of the support $r_m(\beta)$ and the second moment defined as $\bar{r^2}(\beta)=\int dq r^2\rho(r)$ against the value of $\beta$. Both are evaluated numerically. We can see that both quantities behave almost similarly as the function of $\beta$. First, both increase very quickly for small $\beta$ and then both seem to converge toward certain finite values for infinite value of $\beta$
\begin{equation}
\lim_{\beta\rightarrow\infty}r_m\equiv r_{\infty},\hspace{2mm}\lim_{\beta\rightarrow\infty}\bar{r^2}\equiv r^2_{\infty}. 
\label{radius of support at infinite beta}
\end{equation}
In the next section, we shall discuss in more detail the behavior of $\rho(r;\beta)$ and $U(r;\beta)$ for $\beta\rightarrow\infty$.

\begin{figure}[tbp]
\begin{center}
\includegraphics*[width=9cm]{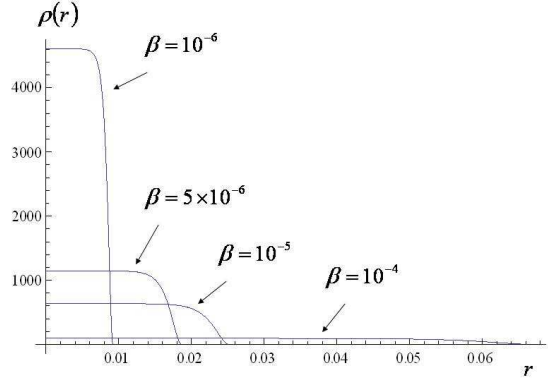}
\end{center}
\caption{The profile of $\rho(r)$ for various small values of $\beta$.}
\label{small beta self-trapped QPD-QP}
\end{figure}

For the opposite limiting case, $\beta\rightarrow 0$, both quantities, $r_m$ and $\bar{r^2}$, are approaching zero. One can thus conclude that the rotationally symmetric self-trapped quantum probability density is converging toward a delta function, 
\begin{equation}
\lim_{\beta\rightarrow 0}\rho({\bf q};\beta)\rightarrow\delta({\bf q}).
\label{delta function at vanishing beta}
\end{equation}
In Fig. \ref{small beta self-trapped QPD-QP}, we confirm the above conclusion numerically by plotting the radial profile of the quantum probability densities for decreasing small values of $\beta=10^{-4},10^{-5},5\times 10^{-6},10^{-6}$. One can also see that for small $\beta$ the quantum probability density is approximately taking a form of a rectangular function. 

Next, let us write the coupled dynamical equations of (\ref{Madelung fluid dynamics}) in polar coordinate, $\{r,\theta\}$, to give us \cite{classical mechanics book}
\begin{eqnarray}
m\Big(\frac{dv_r}{dt}-r\omega^2\Big)=-\partial_rU,\hspace{2mm} \nonumber\\
m\Big(r\frac{d\omega}{dt}+2v_r\omega\Big)=-\partial_{\theta}U,\nonumber\\
\partial_t\rho+\partial_r(\rho v_r)=0.\hspace{12mm}
\label{Madelung's fluid in polar coordinate}
\end{eqnarray}
where $\omega=d\theta/dt$ is the angular velocity and $v_r=dr/dt$. In the last line we have used the fact that our dynamical system is rotationally symmetric. Let us now impose a stationary condition $v_r=0$. First, from the up most equation, the angular velocity is related to the quantum potential as 
\begin{equation}
\omega=\sqrt{\partial_rU/(rm)}.
\label{spinning-stationary condition} 
\end{equation}
Hence, the angular velocity depends only on the distance, $\omega=\omega(r)$. From the middle equation, since $\partial_{\theta}U=0$, one gets $d\omega/dt=0$, such that the angular velocity is constant. Finally, from the lower most equation, one gets $d\rho/dt=\partial_t\rho=0$. Hence, $\rho(r;t)$ is stationary and spinning around its center with a stationary angular velocity field, $\omega(r)$. 

\begin{figure}[tbp]
\begin{center}
\includegraphics*[width=9cm]{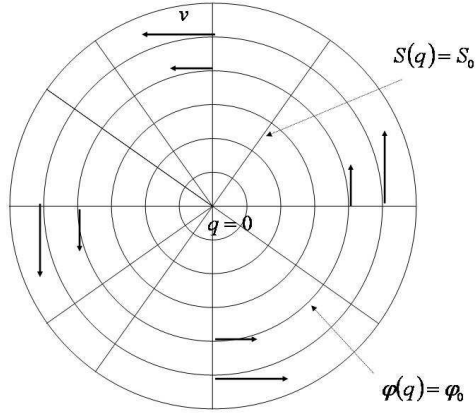}
\end{center}
\caption{The set of concentric circles are the curves of $\varphi({\bf q})=\varphi_0$, whereas the set of lines emanating from ${\bf q}={\bf 0}$ are the curves of $S({\bf q})=S_0$. $\varphi_0$ and $S_0$ are arbitrary constants.}
\label{singular phase}
\end{figure}

Going back to Cartesian coordinate, the rotating velocity vector field can be obtained as 
\begin{equation}
{\bf v}=\omega(r)\times {\bf q}=\{-\omega(r) y,\omega(r) x \}. 
\label{rotating velocity vector field}
\end{equation}
This velocity vector field is divergence-less
\begin{equation}
\partial_{\bf q}\cdot{\bf v}=-\frac{xy}{r}\frac{\partial \omega}{\partial r}+\frac{yx}{r}\frac{\partial\omega}{\partial r}=0.
\label{divergence-less velocity vector field}
\end{equation}
This fact guarantees the existence of a twice differentiable scalar-valued function $\varphi({\bf q})$, a stream function \cite{vector calculus}, such that
\begin{equation}
{\bf v}=\{\partial_y\varphi,-\partial_x\varphi\}.
\label{Poincare theorem: the existence of stream function}
\end{equation}
This means that the tangent of a curve $\varphi({\bf q})=\varphi_0$, where $\varphi_0$ is arbitrary constant, is parallel to the velocity vector ${\bf v}$. The set of curves $\varphi({\bf q})=\varphi_0$ with various values of $\varphi_0$ then make concentric circles with different radius all enclosing ${\bf q}={\bf 0}$. See Fig. \ref{singular phase}.

If one further defines the quantum phase corresponding to this rotational flow as in Eq. (\ref{velocity field-quantum phase}), one will have the following differential equations
\begin{equation}
\partial_xS=m\partial_y\varphi,\hspace{3mm}\partial_yS=-m\partial_x\varphi.
\label{quantum phase rederived}
\end{equation}
From Eqs. (\ref{Poincare theorem: the existence of stream function}) and (\ref{quantum phase rederived}), one gets 
\begin{equation}
\partial_{\bf q}S\cdot\partial_{\bf q}\varphi=m(v_x(-v_y)+v_yv_x)=0.
\label{stream function and stream function are orthogonal}
\end{equation}
Hence all curves $S({\bf q})=S_0$ are orthogonal to all curves $\varphi({\bf q})=\varphi_0$, where $S_0$ and $\varphi_0$ are arbitrary constants. Since the curves $\varphi({\bf q})=\varphi_0$ are a set of concentric circles with different radius depending on the value of $\varphi_0$, then the set of curves $S({\bf q})=S_0$ are line rays emanating from the origin ${\bf q}={\bf 0}$. See Fig. \ref{singular phase}. This means that at ${\bf q}={\bf 0}$, $S({\bf q})$ is not uniquely defined. Hence, at ${\bf q}={\bf 0}$, we have a phase defect or singular point such that Eq. (\ref{irrotational flow}) is not valid. 

Finally, let us notice that Eq. (\ref{quantum phase rederived}) leads to an important fact that $S({\bf q})$ satisfies Laplace equation 
\begin{equation}
\partial_{q}^2S=m\partial_x\partial_y\varphi-m\partial_y\partial_x\varphi=0.
\label{Laplace equation}
\end{equation}
Hence, the quantum phase $S({\bf q})$ of our stationary wave function is a two dimensional harmonic function. On the other hand, the stream function satisfies the following Poisson equation
\begin{equation}
\partial_q^2\varphi=-\partial_{\bf q}\times {\bf v}\neq {\bf 0}. 
\label{Poisson equation}
\end{equation}

Keeping in mind all the above facts about the quantum phase, let us  calculate the average quantum mechanical energy. Writing the wave function in polar form, $\psi=R\exp(iS/\hbar)$, one obtains
\begin{eqnarray}
\langle E\rangle
=\int_{\mathcal{M}} dq\hspace{1mm}R\Big(-\frac{\hbar^2}{2m}\partial_q^2\Big)R
-i\frac{\hbar}{m}\int_{\mathcal{M}} dq\hspace{1mm}R\partial_{\bf q}R\cdot\partial_{\bf q}S\nonumber\\
+\frac{1}{2m}\int_{\mathcal{M}} dq\hspace{1mm}R^2\partial_{\bf q}S\cdot\partial_{\bf q}S
-i\frac{\hbar^2}{2m}\int_{\mathcal{M}} dq\hspace{1mm}\psi^*\partial_q^2S.\hspace{3mm}
\label{average quantum mechanical energy revisited}
\end{eqnarray}
Notice that we have made clear that all the spatial integration is taken over a finite support $\mathcal{M}$ inside which $R({\bf q})\neq 0$. 
The first term on the right hand side is but equal to the average of quantum potential
\begin{equation}
\int_{\mathcal{M}} dq\hspace{1mm}R\Big(-\frac{\hbar^2}{2m}\partial_q^2\Big)R=\int_{\mathcal{M}} dq\hspace{1mm}\rho({\bf q})U({\bf q})=\bar{U}.
\end{equation}
Next, recalling the fact that $R$ depends only on $r$, and Eqs. (\ref{velocity field-quantum phase}), (\ref{rotating velocity vector field}), one has 
\begin{equation}
\partial_{\bf q}R\cdot\partial_{\bf q}S=-m\frac{xy}{r}\omega\partial_rR+m\frac{yx}{r}\omega\partial_rR=0.
\end{equation}
Hence, the second term on the right hand side of Eq. (\ref{average quantum mechanical energy revisited}) is vanishing. Further, using again Eq. (\ref{velocity field-quantum phase}), the third term on the right hand side of Eq. (\ref{average quantum mechanical energy revisited}) can be rewritten as 
\begin{eqnarray}
\frac{1}{2m}\int_{\mathcal{M}} dq\hspace{1mm}R^2\partial_{\bf q}S\cdot\partial_{\bf q}S=\int_{\mathcal{M}} dq\hspace{1mm}\rho\Big(\frac{1}{2}m{\bf v}\cdot {\bf v}\Big) \nonumber\\ 
\equiv \int_{\mathcal{M}} dq\hspace{1mm}\rho({\bf q})K({\bf q})= \bar{K}.\hspace{20mm}
\end{eqnarray}
Here, we have defined a new quantity $K=(1/2)m{\bf v}^2$ whose averaged over $\rho({\bf q})$ is denoted by $\bar{K}$. Namely, $\bar{K}$ can be interpreted as the average kinetic energy of the Madelung fluid. Notice that since ${\bf v}(r)$ depends on $U(r)$, so will $\bar{K}$. We shall show later that $\bar{K}$ depends only on the value of $\beta$, $\bar{K}=\bar{K}(\beta)$. Since $\beta$ depends on $\bar{U}$, one then conclude that $\bar{K}$ depends on $\bar{U}$. Finally, by Eq. (\ref{Laplace equation}), the last term on the right hand side of Eq. (\ref{average quantum mechanical energy revisited}) is vanishing. In total, we thus have the following decomposition of average quantum mechanical energy into the average quantum potential and average kinetic energy
\begin{equation}
\langle E\rangle=\bar{U}+\bar{K}.
\label{internal energy}
\end{equation} 

All the above facts tell us that the self-trapped, rotationally symmetric, spinning yet stationary quantum probability density we developed in this section comprises the class of wave functions that maximizes the Shannon information entropy, given the average quantum mechanical energy of the free particle. \textit{It is thus the most likely wave function of a single free particle in two dimensional space with quantum mechanical energy $\langle E\rangle=\bar{U}+\bar{K}$}. Of course we have to remind ourselves that we have worked in the frame work of Madelung fluid which we assume as the generalization of quantum mechanics allowing rotational flow. 

\section{A Class of Self-Trapped, Non-Spinning Yet Stationary Wave Functions}

Let us discuss the behavior of $\bar{U}$ and $\bar{K}$ of our stationary-spinning wave function as we vary $\beta$. From the stationary condition of Eq. (\ref{spinning-stationary condition}), it is clear that for finite value of $\beta$, the rotating velocity of the Madelung fluid is infinite at the boundary of the support, $\partial\mathcal{M}$. Yet, one should again keep in mind that along this circle of radius $r_m$, $r^2=x^2+y^2=r_m^2$, the quantum probability density is vanishing. For our stationary state, one can then show that the average kinetic energy is finite for non-vanishing value of $\beta$ as follows
\begin{eqnarray}
\bar{K}=\int_{\mathcal{M}} dxdy\hspace{1mm}(1/2)m{\bf v}^2\rho(r)\hspace{40mm}\nonumber\\
=m\pi\int_0^{r_m}dr\hspace{1mm}r^2\partial_rU\rho(r)
=-m\pi\beta^{-1}\int_0^{r_m}dr\hspace{1mm}r^2\partial_r\rho(r)\nonumber\\
=2m\pi\beta^{-1}\int_0^{r_m}dr\hspace{1mm}r\rho(r)=\frac{m}{\beta},\hspace{30mm}\label{kinetic energy versus beta}
\end{eqnarray}
where in the second equality we have used Eq. (\ref{spinning-stationary condition}), in the third equality we have used Eq. (\ref{canonical quantum probability density}) and in the fourth equality we used the partial integration. We have thus an important result which shows that the average kinetic energy $\bar{K}$ depends only on $\beta$, inversely. For non-vanishing $\beta$, assuming a system with finite average quantum mechanical energy $\langle E\rangle$, then the average quantum potential, $\bar{U}=\langle E\rangle-m/\beta$, is also finite. Moreover, given a finite value of average quantum mechanical energy, the definite positivity of $\bar{U}$ leads to the existence of a lower bound for the possible values of $\beta$ as $\beta>m/\langle E\rangle$, such that $r_m$ is also non-vanishing. 

\begin{figure}[tbp]
\begin{center}
\includegraphics*[width=9cm]{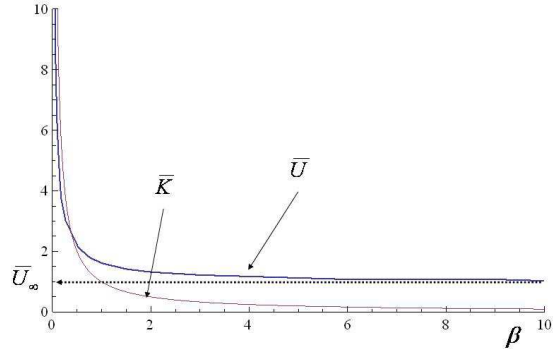}
\end{center}
\caption{The average quantum potential, $\bar{U}$, and average kinetic energy $\bar{K}$ versus $\beta$. $U$ is obtained for the solution of the differential equation (\ref{NPDE for quantum potential}) with fixed boundary conditions, while $\bar{K}$ is obtained from Eq. (\ref{kinetic energy versus beta}).}
\label{energy versus beta}
\end{figure}

Fig. \ref{energy versus beta} shows the numerical plots of the average kinetic energy $\bar{K}$ and the average quantum potential $\bar{U}$ againstf $\beta$ for the stationary-spinning wave functions which satisfy the differential equation (\ref{NPDE for quantum potential}) with fixed boundary conditions and the stationary condition given by Eq. (\ref{kinetic energy versus beta}). One can see that, like $\bar{K}$, $\bar{U}$ decreases as we increase the value of $\beta$. Two limiting cases are of great interest. First is the case when $\beta$ is vanishing, $\beta\rightarrow 0$. As discussed in the previous section, in this case the quantum probability density is approaching a delta function with vanishing support. From Eq. (\ref{kinetic energy versus beta}) one has $\lim_{\beta\rightarrow 0}\bar{K}\rightarrow \infty$. Numerical simulation also shows that the average quantum potential is approaching infinity, $\lim_{\beta\rightarrow 0}\bar{U}\rightarrow\infty$. Hence, this limiting case is not a good model or irrelevant for a single free particle of finite average quantum mechanical energy. 

Let us proceed to discuss the other extreme limiting case of infinite $\beta$, $\beta\rightarrow \infty$. Again, as discussed in the previous section, in this case, the radius of the support $r_m$ and the second moment $\bar{r^2}$ of the quantum probability density are converging toward some certain finite values. This shows that the stationary quantum probability density and its corresponding quantum potential are converging toward some functions
\begin{eqnarray}
\lim_{\beta\rightarrow\infty}\rho({\bf q};\beta)\equiv\rho_{\infty}({\bf q}),\nonumber\\
\lim_{\beta\rightarrow\infty}U({\bf q};\beta)\equiv U_{\infty}({\bf q}).
\label{non-spinning QPD-QP}
\end{eqnarray}
This also tells us that the average quantum potential, $\bar{U}$, is converging toward a finite value, $\lim_{\beta\rightarrow\infty}\bar{U}\equiv\bar{U}_{\infty}$, which is confirmed  by numerical simulation. See Fig. \ref{energy versus beta}. On the other hand, from Eq. (\ref{kinetic energy versus beta}), in this extreme case, the average kinetic energy is vanishing, $\lim_{\beta\rightarrow\infty}\bar{K}\rightarrow 0$. Hence in this limiting case, the average quantum energy is exactly equal to the average quantum potential and thus finite
\begin{equation}
\lim_{\beta\rightarrow\infty}\langle E\rangle=\lim_{\beta\rightarrow\infty}\bar{U}= \bar{U}_{\infty}.
\end{equation}
Moreover, since the average kinetic energy is vanishing, the quantum probability density is no more spinning. This situation is of great interest physically, since we have a lump of self-trapped quantum probability density which is non-spinning yet is still stationary.

\begin{figure}[tbp]
\begin{center}
\includegraphics*[width=9cm]{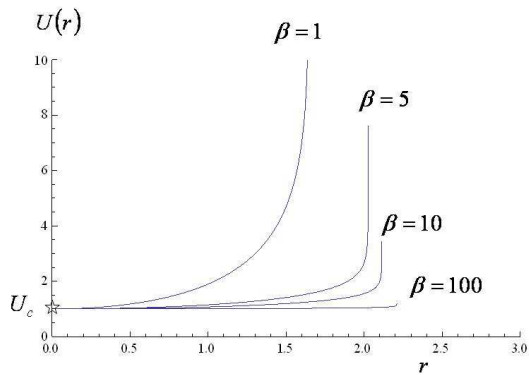}
\end{center}
\caption{The radial profile of quantum potential for various large values of $\beta$.}
\label{the quantum potential at the boundary}
\end{figure}

Let us investigate this last situation in more detail. Fig. \ref{the quantum potential at the boundary} shows the profile of the quantum potential at increasing values of large $\beta$, $\beta=1, 5,10,100$. One can see that the profile of quantum potential is approaching the form of a cylindrical tube with infinite wall, as we increase $\beta$. Hence, for infinite value of $\beta$, the quantum potential is finite and constant inside the support $\mathcal{M}$, and infinite at $\partial{\mathcal{M}}$. Inside the support $\mathcal{M}$, the quantum force is thus vanishing. On the other hand, along the boundary $\partial\mathcal{M}$, the quantum force is infinite. Yet one should keep in mind that along this circle boundary $\partial\mathcal{M}$, the quantum probability density is vanishing. Hence, inside $\mathcal{M}$, both the quantum force and the centrifugal force are vanishing. This ensures that $\rho_{\infty}({\bf q})$ is a non-spinning-stationary solution of the two dimensional Madelung fluid. 

Let us now proceed to discuss the profile of the wave function at the limit $\beta\rightarrow\infty$. As discussed in the previous paragraph, in this limit the quantum potential inside the support is constant. Let us denotes the constant value of $U$ inside $\mathcal{M}$ by $U_{c}$ (See Fig. \ref{the quantum potential at the boundary}). One therefore has 
\begin{eqnarray}
\langle E\rangle=\bar{U}_{\infty}=\int dq\hspace{1mm}\rho_{\infty}({\bf q})U_{c}=U_{c}.
\end{eqnarray}
Next, recalling the definition of quantum potential given in Eq. (\ref{quantum potential}), inside $\mathcal{M}$, one has
\begin{equation}
-\frac{\hbar^2}{2m}\partial_q^2R_{\infty}({\bf q})=U_{c}R_{\infty}({\bf q})=\langle E\rangle R_{\infty}({\bf q}), 
\label{infinite beta wave function 1}
\end{equation}
where we have denoted the quantum amplitude at infinite $\beta$ by $\lim_{\beta\rightarrow\infty}R({\bf q;\beta})\equiv R_{\infty}({\bf q})$. The above differential equation must be subjected to the boundary condition that along the boundary line of the support, $\partial\mathcal{M}$, the quantum amplitude is vanishing. Moreover, since in the limit $\beta\rightarrow\infty$ the phase $S$ is constant inside the support $\mathcal{M}$, multiplying both sides with $\exp(iS/\hbar)$, one gets
\begin{equation}
-\frac{\hbar^2}{2m}\partial_q^2\psi_{\infty}({\bf q})=\langle E\rangle \psi_{\infty}({\bf q}),
\label{infinite beta wave function 2}
\end{equation}
where we have denoted $\psi_{\infty}({\bf q})\equiv R_{\infty}({\bf q})\exp(iS/\hbar)$. 

\begin{figure}[tbp]
\begin{center}
\includegraphics*[width=9cm]{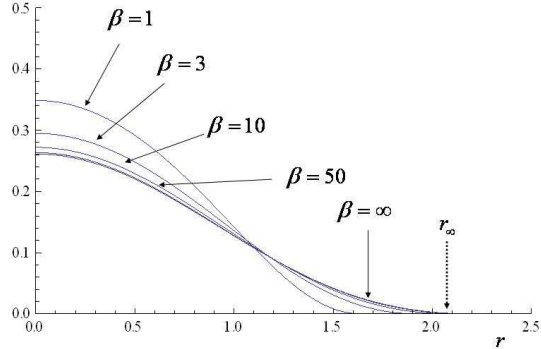}
\end{center}
\caption{$\rho({\bf q};\beta)$ for various values of $\beta$ which solve Eq. (\ref{NPDE for quantum potential}) with fixed boundary conditions and its limiting case for infinite $\beta$, $\rho_{\infty}(r)$, given by Eq. (\ref{infinite beta wave function}).}.
\label{profile of infinite beta wave function}
\end{figure}

Eq. (\ref{infinite beta wave function 2}) with the boundary condition described above is nothing but the stationary (time-independent) Schr\"odinger equation for a particle trapped inside a cylindrical tube classical potential whose bottom is flat and boundary is infinitely high. In polar coordinate, recalling the wave function is rotationally symmetric, one obtains the following differential equation
\begin{equation}
\partial_r^2\psi_{\infty}(r)+\frac{2}{r}\partial_r\psi_{\infty}(r)=-\frac{2m\langle E\rangle}{\hbar^2}\psi_{\infty}(r).
\label{infinite beta wave function 3}
\end{equation}
The solution of which is given by 
\begin{equation}
\psi_{\infty}(r)=AS_c(kr)\exp(iS_0/\hbar),
\label{infinite beta wave function}
\end{equation}
where $S_c(kr)=\sin(kr)/r$ is the Sinc function, $k=\sqrt{2m\langle E\rangle/\hbar^2}$, $A$ is a normalization constant and $S_0$ is an arbitrary phase constant. $k$ is thus a quantity of momentum dimensional. Fig. \ref{profile of infinite beta wave function} shows the quantum probability densities for various increasing large values of $\beta$ obtained by solving Eq. (\ref{NPDE for quantum potential}) with fixed boundary conditions, and its limiting case $\rho_{\infty}(r)$ given by Eq. (\ref{infinite beta wave function}). One can see that as $\beta$ is getting larger, $\rho(r;\beta)$ is converging toward $\rho_{\infty}(r)$. Already at $\beta=50$, $\rho(r;\beta)$ is difficult to be distinguished from $\rho_{\infty}(r)$. Next, from the boundary condition that the quantum probability density must be vanishing along the boundary line of the support, the Sinc function $S_c(kr)$ must reach its first zero point at $r=\lim_{\beta\rightarrow\infty}r_m(\beta)\equiv r_{\infty}$. One therefore has the relation between the average quantum mechanical energy of the single free particle and the radius of the wave function as $kr_{\infty}=\pi$. 

\section{Conclusion and Discussion}

To conclude, we have first assumed the Madelung fluid dynamics as a generalization of Schr\"odinger equation for a single free particle moving in two dimensional space, allowing a rotational flow. We then ask for a class of wave functions whose quantum probability density maximizes Shannon information entropy given the average value of quantum mechanical energy. We showed that there exist a class of solutions in which the quantum probability density is self-trapped, spatially localized with finite support, rotationally symmetric, spinning around its center and yet stationary. Hence it is a quantum vortex in $xy-$space. The stationarity  comes from the balance between the attractive quantum force field of the trapping quantum potential generated by the quantum amplitude and the fictitious repulsive centrifugal force field of the rotating velocity field generated by the quantum phase. 

We then proceeded to show an asymptotic class of solutions in which the quantum probability density is no more spinning and yet still stationary. In this case, the quantum potential takes the form of a cylindrical tube with flat bottom and infinitely high wall. Moreover, we also showed that the wave function turns out to be the lowest eigenstate of the time-independent Schr\"odinger equation for a single particle trapped inside a cylindrical tube classical potential. In this sense, one can consider the class of spinning-stationary solutions of the Madelung fluid dynamics as the generalization of the notion of stationary states of the ordinary quantum mechanics. Yet, one should recall that in constrast to the ordinary quantum mechanics, we did not assume any classical potential to trap the particle. 

Some immediate interesting problems are then ready to be mentioned for future works. It is first interesting to study the case when the initial quantum phase does not satisfy the stationarity condition given in Eq. (\ref{spinning-stationary condition}). In particular, it is instructive to investigate the stability of the stationary solutions with respect to a small perturbation either on the quantum amplitude or quantum phase. In this context, it is interesting to compare the class of localized, stationary and spinning solutions developed in this paper with other localized wave phenomena like soliton etc. The case of three spatial dimension is then of great interest. Moreover, one is tempted to recover the quantization of energy in ordinary quantum mechanics using the formalism developed in this work.

\begin{acknowledgments}

One of the author, A. B., would like to thank Masashi Tachikawa for useful and stimulating discussion. 

\end{acknowledgments}


\begin{thebibliography}{10}

\bibitem{Madelung paper}E. Madelung, Zeits. F. Phys. \textbf{40}, 332 (1926). 

\bibitem{de Broglie papers}L. de Broglie, Compt. Rend. \textbf{183}, 447 (1926); \textbf{184}, 273 (1927); \textbf{185}, 380 (1927). 

\bibitem{Bohm paper}D. Bohm, Phys. Rev. \textbf{85}, 166 (1952); \textbf{85}, 180 (1952); \textbf{89}, 458 (1953). 

\bibitem{Bohm-Hiley's book}D Bohm and B. J. Hiley, \textit{The Undivided Universe: An ontological interpretation of quantum theory} (Routledge, London, 1993).

\bibitem{Holland's book}P. R. Holland, \textit{The Quantum Theory of Motion} (Cambridge University Press, UK, 1993).

\bibitem{Takabayashi paper}Takehiko Takabayashi, Prog. Theor. Phys. \textbf{8}, 143 (1952).  

\bibitem{self-referential dynamics and complexity}N. Kataoka and K. Kaneko, Physica D \textbf{138}, 225 (1999). 

\bibitem{Copenhagen-ontological interpretation}In the so-called pragmatical approach of quantum mechanics, $\rho({\bf q})$ is given meaning as the probability that the particle will be found at ${\bf q}$ if a measurement is performed. On the other hand, in the ontological approach, $\rho({\bf q})$ is argued as the probability that the particle is at ${\bf q}$ regardless of any measurement. See for example \cite{Isham's book,Bohm-Hiley's book,Bell's unspeakable}.

\bibitem{Isham's book}Chris J. Isham, \textit{Lectures On Quantum Theory: Mathematical and Structural Foundation} (Imperial College Press, London, 1995). 

\bibitem{Bell's unspeakable} J. S. Bell, \textit{Speakable and Unspeakable in Quantum Mechanics} (Cambridge University Press, UK, 1987). 

\bibitem{Shannon entropy}C. E. Shannon, Bell Systems Technical Journal \textbf{27}, 379 (1948). 

\bibitem{Cover's book}T. M. Cover and J. A. Thomas, \textit{Elements of Information Theory} (Wiley, USA, 2005).

\bibitem{Jaynes-MEP}E. T. Jaynes, Physical Review \textbf{106}, 620 (1949). 

\bibitem{Shore-Johnson-MEP}J. E. Shore and R. W. Johnson, IEEE Transaction on Information Theory \textbf{IT-26}, 26 (1980).  

\bibitem{Mackey-MEP}M. C. Mackey, Reviews of Modern Physics \textbf{61}, 981 (1989). 

\bibitem{classical mechanics book}V. D. Barger and M. G. Olsson, \textit{Classical Mechanics: A Modern Perspective} (McGraw-Hill Inc., USA, 1995).

\bibitem{vector calculus}George B. Arfken and Hans J. Weber, \textit{Mathematical Methods for Physicits} (Academic Press, San Diego, 1995). 


\end{thebibliography}
\end{document}